\documentclass[prd,onecolumn,nofootinbib]{revtex4}
\usepackage{bm,amsmath,amssymb,graphicx}

\begin{document}

\noindent
Classical and Quantum Gravity {\bf 41} (22), 225001 (2024)\\

\title{Torsional Regularization of Self-Energy and Bare Mass of Electron}

\author{Michael Del Grosso$^{1,2}$}
\author{Nikodem Pop{\l}awski$^2$}
\altaffiliation{NPoplawski@newhaven.edu}

\affiliation{$^{1}$Department of Chemical and Biological Engineering, Rensselaer Polytechnic Institute, Troy, NY, USA}

\affiliation{$^{2}$Department of Mathematics and Physics, University of New Haven, West Haven, CT, USA}

\begin{abstract}
In the presence of spacetime torsion, the momentum components do not commute; therefore, in quantum field theory, summation over the momentum eigenvalues will replace integration over the momentum.
In the Einstein--Cartan theory of gravity, in which torsion is coupled to spin, the separation between the eigenvalues increases with the magnitude of the momentum.
Consequently, this replacement regularizes divergent integrals in Feynman diagrams with loops by turning them into convergent sums.
In this article, we apply torsional regularization to the self-energy of a charged lepton in quantum electrodynamics.
We show that torsion eliminates the ultraviolet divergence of the standard self-energy. 
We also show that the infrared divergence is absent.
In the end, we calculate the finite bare masses of the electron, muon, and tau lepton: $0.4329\,\mbox{MeV}$, $90.95\,\mbox{MeV}$, and $1543\,\mbox{MeV}$, respectively.
These values constitute about $85\%$ of the observed, re-normalized masses.\\ \\
Keywords: torsion, Einstein--Cartan gravity, non-commutative momentum,
regularization,\\ renormalization, electron self-energy, ultraviolet divergence.
\end{abstract}

\maketitle
{\bf 1. Introduction} \\ \\
In quantum electrodynamics, a calculation of the probability amplitude for a physical process will involve higher-order perturbation corrections represented by Feynman diagrams with closed loops of virtual particles \cite{qft,PS}.
At one-loop level, there exist three kinds of diagrams.
The first is vacuum polarization (charge screening): a photon creates a virtual electron--positron pair which annihilates to another photon.
The second is self-energy, which is when an electron emits and reabsorbs a virtual photon.
The third is a vertex, which is when an electron emits a photon, emits a second photon, and then reabsorbs the first \cite{Schw}.
At higher-order levels, all Feynman diagrams are composed from these three fundamental diagrams.

In the four-momentum space, following the principle of superposition, these corrections involve the integration of Feynman propagators. 
These propagators are integrated over unbounded energy and momentum, resulting in integrals whose gauge-invariant parts are logarithmically divergent \cite{qft,PS}.

This non-physical result, called the ultraviolet divergence, is treated by regularization: a mathematical method of modifying the integrals to become finite.
The two most common methods to eliminate this divergence are the Pauli--Villars regularization \cite{PaVi} and the 't Hooft--Veltman dimensional regularization \cite{HoVe}.
When a regulator vanishes, the integral in a four-momentum loop tends to infinity according to a known asymptotic form in the physical limit.
This form can be absorbed by redefining the original (bare) mass, charge, or wave function, leading to a finite (dressed) value that is physically measured in the experiment.
Such redefinition is called re-normalization \cite{qft,ren}.
A bare quantity is divergent, but it is not measurable.

The integrals representing the self-energy and the vertex also have an infrared divergence, which arises from not accounting for soft (low-energy) photons that may exist in a final state \cite{qft,PS}.
When adding soft braking radiation, the infrared divergences in the total cross-section cancel out.
A small mass is given to the photon and then taken to zero in the physical limit to demonstrate this cancellation.

A physical mechanism for regularization was explored in \cite{Rus}, where ultraviolet divergences in quantum field theory might be avoided by curving momentum space.
The idea that momentum space might be curved was first suggested by Born \cite{Born}.
The curvature of momentum space implies the non-commutativity of spacetime coordinates \cite{Snyd}.
This observation has led to the development of non-commutative geometry \cite{Con}, non-commutative field theory \cite{DoNe}, and quantum geometry in which a curved momentum space is dual to a non-commutative spacetime \cite{dual}.

Torsional regularization, based on spacetime torsion \cite{Schr}, provides a physical mechanism that involves non-commutativity of momentum and eliminates ultraviolet-divergent integrals in quantum electrodynamics \cite{toreg}.
The consistency of the conservation law for the total angular momentum of a free Dirac particle in curved spacetime with relativistic quantum mechanics requires torsion \cite{req}.
The most straightforward and most natural theory of gravity with torsion is the Einstein--Cartan theory, where torsion is coupled to spin \cite{EC}.
In the presence of the torsion tensor, the four-momentum components do not commute \cite{toreg}.
A coordinate frame can be chosen such that the momentum components satisfy a commutation relation analogous to that for the angular momentum components \cite{qm}.

In non-commutative momentum space, generated by torsion, integration over the momentum in Feynman diagrams is replaced with summation over the discrete momentum eigenvalues.
Following the Einstein--Cartan equations, the separation between the momentum eigenvalues increases with the magnitude of the momentum.
Consequently, logarithmically divergent integrals in loop diagrams are replaced with convergent sums.
Re-normalization becomes a finite procedure and gives finite forms of the gauge-invariant vacuum polarization tensor, the running coupling, and the bare electric charge of the electron (and other charged particles) \cite{toreg}.

Torsion may therefore physically eliminate infinities arising in quantum field theory.
This elimination would be a remarkable feature of the torsion tensor because torsion, manifesting itself as gravitational repulsion, also removes gravitational singularities in general relativity, replacing the big bang with a non-singular big bounce \cite{avert}.
Instead of forming a singularity, a black hole would create a new universe on the other side of its event horizon \cite{univ}.
Torsion may also explain the cosmic inflation in the early Universe \cite{ApJ}.

Gravitational effects of torsion are strong only at densities much higher than nuclear densities, with a threshold about $10^{50}$ kg/m$^3$ for electrons and $10^{57}$ kg/m$^3$ for neutrons \cite{req}, but torsion can produce noticeable effects in the microscopic regime.
In the presence of torsion, the Dirac equation has a cubic term in the spinor wave function \cite{EC}, which violates the linearity of quantum mechanics at high densities.
The conservation law for the corresponding spin tensor requires fermions to be spatially extended on the order of the Cartan radius, which is about $10^{-27}$ m for the electron and could be eventually measured \cite{nons}.
This extension is related to the ultraviolet-regularizing effect of torsion.

In this article, we apply torsional regularization to the self-energy of a charged lepton.
We review this procedure \cite{toreg} in Section 2.
Next, in Section 3, we review the self-energy of the electron in quantum electrodynamics \cite{PS}.
We show how torsion eliminates the ultraviolet divergence of the self-energy in Section 4, which constitutes the main part of this work.
Lastly, we determine the bare masses of the charged leptons in Section 5.
\[
\]
\noindent
{\bf 2. Torsion and Non-Commutative Momentum} \\ \\
This section reviews torsional regularization, following \cite{toreg}.
The torsion tensor $S^k_{\phantom{k}ij}$ is the antisymmetric part of the affine connection $\Gamma^k_{ij}$ \cite{EC}:
\[
S^k_{\phantom{k}ij}=\frac{1}{2}(\Gamma^k_{ij}-\Gamma^k_{ji}).
\]
In the presence of torsion, the components $p_i$ of the four-dimensional momentum operator satisfy a commutation relation \cite{toreg}:
\[
[p_i,p_j]=2i\hbar S^k_{\phantom{k}ij}p_k.
\]
For the Dirac fields, the spin tensor and thus the torsion tensor are completely antisymmetric \cite{EC}.
In a coordinate frame, in which the pseudovector dual to the torsion tensor has only the temporal component $S^0=-Q/(2\hbar)$, this commutation relation becomes
\[
[p_x,p_y] = iQp_z,\quad [p_y,p_z] = iQp_x,\quad [p_z,p_x] = iQp_y.
\]
Following the analysis in \cite{dim}, the scalar $Q$ is a function of the momentum ${\bf p}$.
To maintain covariance, $Q$ must be a function of the covariant magnitude $p$ of the four-momentum, which is given by $p^2=(p^0)^2-{\bf p}^2$, where $p^0$ is the energy.

According to the Cartan equations, the torsion tensor is a linear function of the spin tensor $s^{\phantom{jk}i}_{jk}$ \cite{EC}:
\[
S^i_{\phantom{i}jk}-S^l_{\phantom{l}jl}\delta^i_k+S^l_{\phantom{l}kl}\delta^i_j=-\frac{1}{2}\kappa s^{\phantom{jk}i}_{jk}.
\]
We use $\hbar=c=1$.
The spin tensor describes the spin density (spin per volume).
Because of the de Broglie formula, which relates the momentum of a particle to the inverse of the wavelength of the corresponding wave \cite{qm}, and the volume occupied by the particle on the order of the cube of the wavelength, the spin tensor is a cubic function of $p$ \cite{toreg}.
The quantity $Q$ has the dimension of the torsion tensor and thus of the spin tensor multiplied by $\kappa$.
Because $\kappa=8\pi G\sim M_\textrm{P}^{-2}$, where $M_\textrm{P}$ is the Planck mass, $Q$ has a form
\[
Q=Up^3,
\]
where a constant $U$, which can be taken positive without loss of generality, is on the order of $1/M_\textrm{P}^2$.
The cyclic commutation relations for the momentum can be written as 
\begin{equation}
[n_x,n_y]=in_z,\quad [n_y,n_z]=in_x,\quad [n_z,n_x]=in_y 
\label{commute}
\end{equation}
for the spatial components $n_i$ of a vector
\begin{equation}
{\bf n}=\frac{{\bf p}}{Q}.
\label{normal}
\end{equation}

The relations (\ref{commute}) are analogous to those for the angular momentum.
Consequently, the eigenvalues of ${\bf n}$ are given by $n=|{\bf n}|=\sqrt{j(j+1)}$ and $n_z=m$, where an integer $j\ge0$ is the orbital quantum number and an integer $m\in[-j,j]$ is the magnetic quantum number \cite{qm}.

Torsional regularization \cite{toreg} replaces integration over the momentum in Feynman diagrams with summation over the momentum eigenvalues related to the eigenvalues of ${\bf n}$:
\begin{equation}
\iiint dn_x dn_y dn_z\, f(n)\rightarrow 4\pi \sum_{\textrm{eigenstates}} f(n)\,|n_z|=4\pi\sum_{j=1}^\infty\sum_{m=-j}^j f(n)\,|m|=4\pi \sum_{j=1}^\infty f(n)\,j(j+1)=4\pi \sum_{j=1}^\infty f(n)\,n^2,
\label{pres}
\end{equation}
where $f(n)$ is an arbitrary scalar function of $n$.
This replacement follows from the commutation relations (\ref{commute}) \cite{toreg}.
According to this prescription, such a function is multiplied by the absolute value of the commutator of two integration variables, $|[n_x,n_y]|=|n_z|$, and integration over continuous variables $n_x$, $n_y$, $n_z$ is replaced with summation over the eigenvalues of ${\bf n}$.
For $j=0$, $m=0$ and thus $n_z=0$, which does not contribute to the sum in (\ref{pres}).\footnote{
The prescription proposed in \cite{toreg} is based on the correspondence between the classical and quantum partition functions:
\[
\frac{1}{2\pi\hbar}\iint dq\,dp \leftrightarrow \sum_{\textrm{eigenstates}},\quad \iint dq\,dp\, f(H(q,p)) \leftrightarrow 2\pi\sum_{\textrm{eigenstates}} f(E)\,|[q,p]|,
\label{partition}
\]
where $f(x)=e^{-x/kT}$.
The quantum commutation relation between the integration variables $q$ and $p$ leads to a discrete spectrum of energy eigenstates.
Therefore, integration over a continuous phase space in a classical partition function is replaced with summation over energy eigenstates in the corresponding quantum partition function.
}

A Feynman diagram with a loop involves calculating an integral in the four-momentum space that has a form $\int d^4l/(l^2-\Delta+i\epsilon)^s$, where $\Delta>0$ does not depend on the four-momentum $l^i$, $s$ is an integer, and $\epsilon\to 0^{+}$ \cite{qft,PS}.
Applying the Wick rotation, in which the temporal component of the four-momentum $l^0$ is replaced with $il^0_\textrm{E}$ and thus $l^2_\textrm{E}=(l^0_\textrm{E})^2+{\bf l}^2_\textrm{E}$, turns the integration in spacetime with the Lorentzian metric signature to the integration in the four-dimensional Euclidean space.
The integral becomes $i(-1)^s\int d^4l_\textrm{E}/(l^2_\textrm{E}+\Delta)^s=i(-1)^s\int dl_\textrm{E}^0 d{\bf l}/(l^2_\textrm{E}+\Delta)^s$, where $d{\bf l}=dl_x dl_y dl_z$.
Omitting the subscript E and using $Q=Ul^3$ with $U>0$ and (\ref{normal}) gives 
\begin{equation}
l^2=(l^0)^2+{\bf l}^2=(l^0)^2+U^2 n^2 l^6.
\label{condit}
\end{equation}
The integration over $l^0$ can be replaced with the integration over $l$:
\[
dl^0=dl \frac{dl^0}{dl}=dl \frac{1-3U^2 n^2 l^4}{(1-U^2 n^2 l^4)^{1/2}}.
\]
The integration over ${\bf l}$ can be replaced with the integration over ${\bf n}={\bf l}/Q$ and then changed to the summation over $j$ in the non-commutative momentum space:
\begin{equation}
\iiint dl_x dl_y dl_z\, f({\bf l}^2)\rightarrow \iiint dn_x dn_y dn_z J\,f(Q^2 n^2)\rightarrow 4\pi \sum_{j=1}^\infty J\,f(Q^2 n^2)\,n^2,
\label{prescr}
\end{equation}
where $J=\partial(l_x,l_y,l_z)/\partial(n_x,n_y,n_z)$ is the Jacobian of the transformation from ${\bf l}$ to ${\bf n}$.

Differentiating (\ref{condit}) with respect to $n_x$, using $n^2=n_x^2+n_y^2+n_z^2$, gives $2l(\partial l/\partial n_x)=6U^2 n^2 l^5 (\partial l/\partial n_x)+2U^2 l^6 n_x$, which is equivalent to
\[
\frac{\partial l}{\partial n_x}=\frac{U^2 l^5 n_x}{1-3U^2 n^2 l^4}.
\]
The transformation derivatives are thus
\begin{eqnarray}
& & \frac{\partial l_x}{\partial n_x}=\frac{\partial(Qn_x)}{\partial n_x}=Q+3Un_x l^2 \frac{\partial l}{\partial n_x}=\frac{Q}{1-3U^2 n^2 l^4}[1-3U^2 l^4 (n_y^2 + n_z^2)], \nonumber \\
& & \frac{\partial l_x}{\partial n_y}=\frac{\partial(Qn_x)}{\partial n_y}=3Un_x l^2 \frac{\partial l}{\partial n_y}=\frac{Q}{1-3U^2 n^2 l^4}(3U^2 l^4 n_x n_y), \nonumber
\end{eqnarray}
and similarly for other components.
They give the Jacobian:
\[
J=\mbox{det}\left( \begin{array}{ccc}
\partial l_x/\partial n_x & \partial l_x/\partial n_y & \partial l_x/\partial n_z \\
\partial l_y/\partial n_x & \partial l_y/\partial n_y & \partial l_y/\partial n_z \\
\partial l_z/\partial n_x & \partial l_z/\partial n_y & \partial l_z/\partial n_z \end{array} \right)=\frac{Q^3}{1-3U^2 n^2 l^4}.
\]

Consequently, integrating over the Euclidean four-momentum $l^i$ and using the prescription (\ref{prescr}) gives
\begin{eqnarray}
& & \int dl^0 d{\bf l}=\int dl \frac{dl^0}{dl} J\,d{\bf n}=\int dl\,d{\bf n}\frac{Q^3}{(1-U^2 n^2 l^4)^{1/2}}=2\int_0^{\sqrt{1/(Un)}}dl\,d{\bf n}\frac{U^3 l^9}{(1-U^2 n^2 l^4)^{1/2}} \nonumber \\
& & \rightarrow 8\pi \sum_{j=1}^\infty\int_0^{\sqrt{1/(Un)}}dl \frac{U^3 l^9}{(1-U^2 n^2 l^4)^{1/2}} n^2, \nonumber
\end{eqnarray}
with $n=\sqrt{j(j+1)}$.
The integral $\int dl^0 d{\bf l}/(l^2+\Delta)^s$ in the presence of the Einstein--Cartan torsion is therefore \cite{toreg}
\begin{widetext}
\begin{eqnarray}
& & \int dl^0 d{\bf l}\frac{1}{(l^2+\Delta)^s}\rightarrow 8\pi \sum_{j=1}^\infty\int_0^{\sqrt{1/(Un)}}dl \frac{U^3 l^9}{(1-U^2 n^2 l^4)^{1/2}} n^2\frac{1}{(l^2+\Delta)^s}=8\pi \sum_{j=1}^\infty\int_0^1 d\xi \frac{U^3 \xi^9 n^2 (Un)^{-5}}{(1-\xi^4)^{1/2}[\xi^2/(Un)+\Delta]^s} \nonumber \\
& & =\sum_{j=1}^\infty\int_0^1 d\xi \frac{8\pi U^{-2} \xi^9 n^{-3} (Un)^s}{(1-\xi^4)^{1/2}(\xi^2+U\Delta n)^s}=\sum_{j=1}^\infty\int_0^1 d\zeta \frac{4\pi U^{s-2} \zeta^4 n^{s-3}}{(1-\zeta^2)^{1/2}(\zeta+U\Delta n)^s}=4\pi \sum_{j=1}^\infty\int_0^{\pi/2} d\phi \frac{U^{s-2} \sin^4\phi\,n^{s-3}}{(\sin\phi+U\Delta n)^s} \nonumber \\
& & =4\pi \sum_{j=1}^\infty\int_0^{\pi/2} d\phi \frac{U^{s-2} \sin^4\phi\,(j(j+1))^{(s-3)/2}}{[\sin\phi+U\Delta (j(j+1))^{1/2}]^s},
\label{sumint}
\end{eqnarray}
\end{widetext}
using a series of substitutions: $Unl^2=\xi^2=\zeta=\sin\phi$.
For $U<0$, $U$ in the sum-integral (\ref{sumint}) is replaced with $|U|$.
At large values of $j$, the sum-integral (\ref{sumint}) behaves as $\sim\sum_{j=1}^\infty j^{-3}$ for any $s$, so it converges, regularizing the original integral.

In the limit of continuous momentum space, $U\rightarrow 0$, the separation between adjacent values of $j$ has no effect on the integral, so the summation over $j$ can be replaced with the integration over $y$, where $y=\sqrt{j(j+1)}$.
For $s=3$, this limit is equal to the original, finite value of the integral.
For $s=2$, the integral in this limit diverges as $\int d^4 l/l^4\sim\int dl/l\sim\ln(U)$, equivalently to the Pauli--Villars regularization \cite{PaVi} with the mass $\Lambda\to\infty$ of a fictitious heavy photon given by $U=1/\Lambda^2$ \cite{toreg}, and to the dimensional regularization \cite{HoVe} with the number of dimensions $n\to 4^+$ given by $1/(n-4)+(1/2)[\ln(4\pi)-\gamma]=\ln\Lambda$, where $\gamma$ is the Euler--Mascheroni constant.
\[
\]
{\bf 3. Electron Self-Energy} \\ \\
This section reviews the self-energy of the electron in quantum electrodynamics, following \cite{PS}.
The Feynman propagator for an electron with bare mass $m_0$ and four-momentum $p_\mu$ is given by
\[
\mathcal{G}=\frac{i(\not p + m_{0})}{(p^{2} - m_{0}^{2} + i\epsilon)},
\]
where $\not p=\gamma^\mu p_\mu$ and $\gamma^\mu$ are the Dirac matrices.
The propagator in the second-order in $e$ is given by
\[
\mathcal{G}\Sigma\mathcal{G}=\frac{i(\not p + m_{0})}{(p^{2} - m_{0}^{2} + i\epsilon)}[-i\Sigma_{2}(p)]\frac{i(\not p + m_{0})}{(p^{2} - m_{0}^{2} + i\epsilon)},
\]
where $\Sigma_{2}$ represents the electron self-energy at one-loop level:
\[
-i\Sigma_{2}(p)=(-ie)^{2}\int\frac{d^{4}k}{(2\pi)^{4}}\gamma^{\mu}\frac{i(\not k + m_{0})}{k^{2}-m_{0}^{2}+i\epsilon}\gamma_{\mu}\frac{-i}{(p-k)^{2}-\mu^{2}+i\epsilon}.
\]
Infrared divergence is regularized by adding a small photon mass $\mu$.
This propagator describes an electron that emits a photon and reabsorbs it.

Introducing a Feynman parameter $x$ combines two denominators into one:
\[
\frac{1}{k^{2}-m_{0}^{2}+i\epsilon}\frac{1}{(p-k)^{2}-\mu^{2}+i\epsilon}=\int_{0}^{1}\frac{1}{[(k^{2}-2xk\cdot p+xp^{2}-x\mu^{2}-(1-x)m_{0}^{2}+i\epsilon)]^{2}}dx.
\]
Consequently, one obtains
\[
i\Sigma_2(p)=-e^2\int_{0}^{1}dx\int\frac{d^{4}l}{(2\pi)^{4}}\frac{-2x{\not p} + 4m_{0}}{[l^{2}-\Delta+i\epsilon]^{2}},
\]
where
\[
\Delta = -x(1-x)p^{2}+x\mu^2+(1-x)m_{0}^{2}.
\]
Applying the Wick rotation gives an integral in the four-dimensional Euclidean space:
\[
\Sigma_2(p)=e^2\int_{0}^{1}dx\int\frac{d^{4}l_{E}}{(2\pi)^{4}}\frac{-2x{\not p} + 4m_{0}}{[l_{E}^{2}+\Delta]^2}.
\]

The full propagator of an electron is equal to the sum of terms that involve powers of one-particle-irreducible diagrams:
\[
\mathcal{G}+\mathcal{G}\Sigma\mathcal{G}+\mathcal{G}\Sigma\mathcal{G}\Sigma\mathcal{G}+\dots=\frac{i}{\not p-m_0-\Sigma(\not p)}.
\]
The physical (observed) mass $m$ is located at a pole of the above denominator that is a solution of
\[
[\not p - m_{0}-\Sigma(\not p)]|_{\not p=m}=0.
\]
The full propagator has a simple pole shifted from $m_0$ by the self-energy $\Sigma({\not p})$.
Near the pole, the denominator is equal to
$({\not p} - m)(1-d\Sigma/d{\not p})|_{\not p=m}$,
which gives the re-normalization constant of the electron wave function:
$Z_2^{-1}=(1-d\Sigma/d{\not p})|_{\not p=m}$.
Therefore, the physical mass is shifted from the bare mass in the second order in $e$ by
\[
\delta m=m-m_0=\Sigma_2(\not p=m)\approx \Sigma_2(\not p=m_0).
\]
Consequently, one obtains
\begin{equation}
\delta m=2m_0 e^2\int_{0}^{1}dx\int\frac{d^{4}l_{E}}{(2\pi)^{4}}\frac{2-x}{[l_{E}^{2}+\Delta]^2},
\label{self}
\end{equation}
where
\begin{equation}
\Delta = (1-x)^{2}m_{0}^{2}+x\mu^2.
\label{delta}
\end{equation}
\[
\]
{\bf 4. Torsional Regularization of Ultraviolet Divergence} \\ \\
Now, we proceed to the main calculation of this work.

In the presence of torsion, integration over the momentum is replaced with summation over the momentum eigenvalues according to (\ref{sumint}):
\[
\int\frac{d^4l_{E}}{(l^{2}_{E}+\Delta)^s}\rightarrow 4\pi U^{s-2}\sum_{l=1}^{\infty}\int_0^{\pi/2} d\phi \frac{\sin^4\phi\,[l(l+1)]^{(s-3)/2}}{[\sin\phi+U\Delta\sqrt{l(l+1)}]^s}.
\]
For $s=2$, this replacement gives
\begin{equation}
\int\frac{d^4l_{E}}{(l^{2}_{E}+\Delta)^2}\rightarrow 4\pi\sum_{l=1}^\infty \int_0^{\pi/2} d\phi \frac{\sin^4\phi\,[l(l+1)]^{-1/2}}{[\sin\phi+U\Delta\sqrt{l(l+1)}]^2}.
\label{sumintse}
\end{equation}
With the prescription (\ref{sumintse}), the mass shift (\ref{self}) becomes
\begin{equation}
\delta m=\frac{8\pi m_{0}e^{2}}{(2\pi)^{4}}\sum_{l=1}^{\infty}\int_{0}^{\pi/2}d \phi \int_{0}^{1}dx\frac{\sin^{4}\phi(2-x)[l(l+1)]^{-1/2}}{[\sin \phi + U\Delta \sqrt{l(l+1)}]^{2}}.
\label{mass}
\end{equation}

Since $U\sim M^{-2}_\textrm{P}$, $U\Delta$ is on the order of $(m_0/M_\textrm{P})^2\ll 1$.
In this limit, the separation between adjacent values of $l$ does not affect significantly the integrand and thus the summation over $l$ in (\ref{mass}) can be replaced (to make calculations simpler) with the integration over $n$, where $n=\sqrt{l(l+1)}$.
Consequently, we obtain
\[
\delta m=\frac{8\pi m_{0}e^{2}}{(2\pi)^{4}}
\int_{0}^{\pi/2}d \phi \int_{0}^{1}dx\int_{\sqrt{2}}^\infty dn\frac{\sin^{4}\phi(2-x)n^{-1}}{[\sin \phi + U\Delta n]^{2}}.
\]
Substituting from $n$ to a new variable $w=U\Delta n$ gives
\begin{eqnarray}
& & \delta m=\frac{8\pi m_{0}e^{2}}{(2\pi)^{4}}
\int_{0}^{\pi/2}d \phi\, \sin^{4}\phi \int_{0}^{1}dx(2-x)\int_{\sqrt{2}U\Delta}^\infty \frac{dw}{w[\sin \phi + w]^{2}} \nonumber \\
& & =\frac{2m_{0}\alpha}{\pi^{2}}
\int_{0}^{1}dx(2-x)\int_{0}^{\pi/2}d \phi\Bigl[-\frac{\sin^3\phi}{\sin\phi+\sqrt{2}U\Delta}+\sin^{2}\phi\ln\Bigl(\frac{\sin\phi}{\sqrt{2}U\Delta}+1\Bigr)\Bigr],
\label{shift}
\end{eqnarray}
where $\alpha=e^2/4\pi$ is the fine structure constant.
This integral is convergent as long as $U\neq 0$, that is, in the presence of torsion, and as long as $\Delta\neq0$.
\[
\]
{\bf 5. Calculation and Discussion} \\ \\
Because of torsion, the constant $U$ is different from zero, which eliminates the ultraviolet divergence \cite{toreg}.
Following Section 2, this constant is on the order of the squared inverse of the Planck mass and its exact value could be derived by solving the Dirac equation in curved spacetime \cite{nons} instead of the estimation based on the de Broglie formula.
For simplicity, we choose
\begin{equation}
U=\frac{1}{M_\textrm{P}^2}.
\label{photon}
\end{equation}
As long as $\mu>0$, $\Delta$ in (\ref{delta}) is different from zero for the entire range of $x\in[0,1]$.
Because $U\Delta\ll 1$, the integral (\ref{shift}) can be approximated as
\begin{eqnarray}
& & \delta m=\frac{2m_{0}\alpha}{\pi^{2}}
\int_{0}^{1}dx(2-x)\int_{0}^{\pi/2}d \phi\Bigl[-\frac{\sin^3\phi}{\sin\phi}+\sin^{2}\phi\ln\Bigl(\frac{\sin\phi}{\sqrt{2}U\Delta}\Bigr)\Bigr] \nonumber \\
& & =\frac{2m_{0}\alpha}{\pi^{2}}
\int_{0}^{1}dx(2-x)\int_{0}^{\pi/2}d \phi\Bigl[-\sin^2\phi-\sin^{2}\phi\ln\Bigl(\sqrt{2}Um_{0}^{2}\frac{\Delta}{m_{0}^{2}}\Bigr)+\sin^{2}\phi\ln(\sin\phi)\Bigr] \nonumber \\
& & =\frac{2m_{0}\alpha}{\pi^{2}}
\int_{0}^{1}dx(2-x)\Biggl[\int_{0}^{\pi/2}d \phi\sin^2\phi\Bigl[-1-\ln(\sqrt{2}U m_{0}^{2})-\ln\Bigl(\frac{\Delta}{m_{0}^{2}}\Bigr)\Bigr]+\int_{0}^{\pi/2}d \phi\sin^2\phi\ln(\sin\phi)\Biggr], \nonumber
\end{eqnarray}
where $\Delta$ is given by (\ref{delta}).
After integrating over $\phi$, we obtain
\[
\delta m=\frac{2m_{0}\alpha}{\pi^{2}}
\int_{0}^{1}dx(2-x)\Biggl[\frac{\pi}{4}\Bigl[(-1-\ln(\sqrt{2}Um_{0}^{2})-\ln\Bigl(\frac{\Delta}{m_{0}^{2}}\Bigr)\Bigr]+N\Biggr],
\]
where
\[
N=\int_0^{\pi/2}\sin^2\phi\ln(\sin\phi)\,d\phi\approx-0.151697.
\]
Consequently, the mass shift is equal to
\begin{eqnarray}
& & \delta m=\frac{2m_{0}\alpha}{\pi^{2}}\frac{3}{2}\Bigl[\frac{\pi}{4}\bigl(-1-\ln(\sqrt{2}Um_{0}^{2})\bigr)+N\Bigr]-\frac{2m_{0}\alpha}{\pi^{2}}\int_{0}^{1}dx(2-x)\frac{\pi}{4}\ln\Bigl(\frac{\Delta}{m_{0}^{2}}\Bigr) \nonumber \\
& & =\frac{3m_{0}\alpha}{4\pi}\Bigl(\frac{4N}{\pi}-1-\ln(\sqrt{2}Um_{0}^{2})\Bigr)-\frac{m_{0}\alpha}{2\pi}\int_{0}^{1}dx(2-x)\ln\Bigl((1-x)^{2}+x\frac{\mu^{2}}{m_{0}^{2}}\Bigr).
\label{final}
\end{eqnarray}

With the substitution of $U$ from (\ref{photon}) and taking a limit $\mu\to0$, we use the relation (\ref{final}) to determine the observed mass $m=m_0+\delta m$ as a function of the bare mass $m_0$.
Conversely, we obtain the bare mass from the observed mass.
The integral in (\ref{final}) converges in this limit, so the infrared divergence of a lepton self-energy is absent.
Following \cite{order}, infrared divergences are canceled to all orders.
For the electron, $m=0.510999\,\mbox{MeV}$ gives $m_0=0.432928\,\mbox{MeV}$.
For the muon, $m=105.658\,\mbox{MeV}$ gives $m_0=90.9514\,\mbox{MeV}$.
For the tau lepton, $m=1776.86\,\mbox{MeV}$ gives $m_0=1542.63\,\mbox{MeV}$.

Our results depend on the constant $U$ estimated in (\ref{photon}), but they are not significantly affected by changing its exact value.
For the electron, increasing $U$ by a factor of 10 gives $m_0=0.434408\,\mbox{MeV}$ and decreasing it by the same factor gives $m_0=0.431457\,\mbox{MeV}$.
We conclude that the bare masses of charged leptons constitute about $85\%$ of their observed, re-normalized masses.
To compare, the bare electric charge is about $22\%$ higher than the observed, re-normalized charge \cite{toreg}.

These numbers might change slightly when the contributions from the W and Z bosons are included.
Because the propagators for these bosons and for gluons are similar to the photon propagator, we think that torsional regularization, replacing the integration over continuous momentum in Feynman diagrams with the summation over discrete momentum, can be extended to non-Abelian gauge theories, turning divergent integrals into convergent sums.
The only issue is that torsion may couple to the W and Z fields because they are massive; it does not couple to photons because of the gauge invariance resulting from the massless photon \cite{req}.
The effects of such a possible coupling should be investigated.
\[
\]
{\bf Acknowledgments} \\ \\
I am grateful to Francisco Guedes and my Parents, Bo\.{z}enna Pop{\l}awska and Janusz Pop{\l}awski, for inspiring this research - N P.


\begin{thebibliography}{}
\bibitem{qft} R. P. Feynman, Phys. Rev. {\bf 76}, 769 (1949); J. D. Bjorken and S. D. Drell, {\it Relativistic Quantum Fields} (McGraw-Hill, 1965); C. Nash, {\it Relativistic Quantum Fields} (Academic, 1978); F. Mandl and G. Shaw, {\it Quantum Field Theory} (Wiley, 1993); W. Greiner and J. Reinhardt, {\it Quantum Electrodynamics} (Springer, 1994); M. Maggiore, {\it A Modern Introduction to Quantum Field Theory} (Oxford University Press, 2005); C. Itzykson and J.-B. Zuber, {\it Quantum Field Theory} (Dover, 2006); K. Huang, {\it Quantum Field Theory: From Operators to Path Integrals} (Wiley, 2010).
\bibitem{PS} M. E. Peskin and D. V. Schroeder, {\it An Introduction to Quantum Field Theory} (Addison-Wesley, 1995).
\bibitem{Schw} J. Schwinger, Phys. Rev. {\bf 75}, 651 (1949); Phys. Rev. {\bf  82}, 664 (1951).
\bibitem{PaVi} W. Pauli and F. Villars, Rev. Mod. Phys. {\bf 21}, 434 (1949).
\bibitem{HoVe} G. 't Hooft and M. Veltman, Nucl. Phys. B {\bf 44}, 189 (1972).
\bibitem{ren} F. J. Dyson, Phys. Rev. {\bf 75}, 486 (1949); M. Gell-Mann and F. E. Low, Phys. Rev. {\bf 95}, 1300 (1954).
\bibitem{Rus} Y. A. Gol'fand, Sov. Phys. J. Exp. Theor. Phys. {\bf 10}, 356 (1960); V. G. Kadyshevskii, Sov. Phys. J. Exp. Theor. Phys. {\bf 14}, 1340 (1962); Y. A. Gol'fand, Sov. Phys. J. Exp. Theor. Phys. {\bf 16}, 184 (1963); I. P. Volobuev, V. G. Kadyshevskii, M. D. Mateev, and R. M. Mir-Kasimov, Theor. Math. Phys. {\bf 40}, 800 (1979); V. G. Kadyshevsky and M. D. Mateev, Nuovo Cim. A {\bf 87}, 324 (1985).
\bibitem{Born} M. Born, Proc. R. Soc. A {\bf 165}, 291 (1938); Rev. Mod. Phys. {\bf 21}, 463 (1949).
\bibitem{Snyd} H. S. Snyder, Phys. Rev. {\bf 71}, 38 (1947); C. N. Yang, Phys. Rev. {\bf 72}, 874 (1947).
\bibitem{Con} A. Connes, {\it Noncommutative Geometry} (Academic, 1994).
\bibitem{DoNe} M. R. Douglas and N. A. Nekrasov, Rev. Mod. Phys. {\bf 73}, 977 (2001); M. Arzano, Phys. Rev. D {\bf 83}, 025025 (2011).
\bibitem{dual} S. Majid, Lect. Notes Phys. {\bf 541}, 227 (2000).
\bibitem{Schr} E. Schr\"{o}dinger, {\it Space-Time Structure} (Cambridge University Press, 1954).
\bibitem{toreg} N. Pop{\l}awski, Found. Phys. {\bf 50}, 900 (2020); arXiv:1807.07068 (2018).
\bibitem{req} F. W. Hehl, P. von der Heyde, G. D. Kerlick, and J. M. Nester, Rev. Mod. Phys. {\bf 48}, 393 (1976).
\bibitem{EC} T. W. B. Kibble, J. Math. Phys. {\bf 2}, 212 (1961); D. W. Sciama, in {\em Recent Developments in General Relativity}, p. 415 (Pergamon, 1962); Rev. Mod. Phys. {\bf 36}, 463 (1964); Rev. Mod. Phys. {\bf 36}, 1103 (1964); F. W. Hehl and B. K. Datta, J. Math. Phys. {\bf 12}, 1334 (1971); E. A. Lord, {\em Tensors, Relativity and Cosmology} (McGraw-Hill, 1976); V. de Sabbata and M. Gasperini, {\it Introduction to Gravitation} (World Scientific, 1985); K. Nomura, T. Shirafuji, and K. Hayashi, Prog. Theor. Phys. {\bf 86}, 1239 (1991); V. de Sabbata and C. Sivaram, {\it Spin and Torsion in Gravitation} (World Scientific, 1994); N. Pop{\l}awski, {\it Classical Physics: Spacetime and Fields}, arXiv:0911.0334 (2024); F. R. Benard Guedes and N. J. Pop{\l}awski, Class. Quantum Grav. {\bf 41}, 065011 (2024).
\bibitem{qm} P. A. M. Dirac, {\it The Principles of Quantum Mechanics} (Oxford University Press, 1930); L. D. Landau and E. M. Lifshitz, {\it Quantum Mechanics: Non-Relativistic Theory} (Pergamon, 1977); J. J. Sakurai, {\it Modern Quantum Mechanics} (Addison-Wesley, 1994).
\bibitem{avert} F. W. Hehl, P. von der Heyde, and G. D. Kerlick, Phys. Rev. D {\bf 10}, 1066 (1974); B. Kuchowicz, Gen. Relativ. Gravit. {\bf 9}, 511 (1978); M. Gasperini, Phys. Rev. Lett. {\bf 56}, 2873 (1986).
\bibitem{univ} N. J. Pop{\l}awski, Phys. Lett. B {\bf 694}, 181 (2010); Phys. Lett. B {\bf 701}, 672 (2011); Gen. Relativ. Gravit. {\bf 44}, 1007 (2012); N. Pop{\l}awski, Phys. Rev. D {\bf 85}, 107502 (2012); G. Unger and N. Pop{\l}awski, Astrophys. J. {\bf 870}, 78 (2019); J. L. Cubero and N. J. Pop{\l}awski, Class. Quantum Grav. {\bf 37}, 025011 (2020); N. Pop{\l}awski, Zh. Eksp. Teor. Fiz. {\bf 159}, 448 (2021); J. Exp. Theor. Phys. {\bf 132}, 374 (2021); in: {\em Proceedings of the Sixteenth Marcel Grossmann Meeting on General Relativity}, ed. R. Ruffini and G. Vereshchagin, part B, p. 1327 (World Scientific, 2023); in: {\em Regular Black Holes. Towards a New Paradigm of Gravitational Collapse}, ed. C. Bambi, p. 485 (Springer, 2023).
\bibitem{ApJ} N. Pop{\l}awski, Astrophys. J. {\bf 832}, 96 (2016); S. Desai and N. J. Pop{\l}awski, Phys. Lett. B {\bf 755}, 183 (2016).
\bibitem{nons} N. J. Pop{\l}awski, Phys. Lett. B {\bf 690}, 73 (2010); Phys. Lett. B {\bf 727}, 575 (2013).
\bibitem{dim} N. Sasakura, J. High Energy Phys. {\bf 05}, 015 (2000); Y. Sasai and N. Sasakura, J. High Energy Phys. {\bf 06}, 013 (2009).
\bibitem{order} J. D. More and A. Misra, Phys. Rev. D {\bf 89}, 105021 (2014).
\end{thebibliography}
\end{document}